\documentclass[final,times,onecolumn]{elsarticle}
\biboptions{sort&compress}
\usepackage[margin=1in]{geometry}
\usepackage[utf8]{inputenc}
\usepackage{bm}
\usepackage[dvipsnames]{xcolor}
\usepackage{multicol}
\usepackage{multirow}
\usepackage[normalem]{ulem}
\usepackage{subfig}
\usepackage{graphicx}
\usepackage{amssymb}
\usepackage{placeins}
\usepackage{amsmath}
\usepackage{slashed}
\usepackage{hyperref}
\allowdisplaybreaks

\newcommand{\sa}{{\scalebox{0.75}{$\sphericalangle$}}}

\begin{document}

\title{Transverse-spin dependent energy-energy correlators in proton-proton collisions within the dihadron fragmentation framework}

\author[label1,label2,label3]{Zhong-Bo Kang} \ead{zkang@physics.ucla.edu}
\author[label4]{Andreas Metz} \ead{metza@temple.edu}
\author[label5]{Daniel Pitonyak} \ead{pitonyak@lvc.edu}
\author[label1,label2]{Congyue Zhang} \ead{maxzhang2002@g.ucla.edu}

\address[label1]{Department of Physics and Astronomy, University of California, Los Angeles, CA 90095, USA}
\address[label2]{Mani L. Bhaumik Institute for Theoretical Physics, University of California, Los Angeles, CA 90095, USA}
\address[label3]{Center for Frontiers in Nuclear Science, Stony Brook University, Stony Brook, NY 11794, USA}
\address[label4]{Department of Physics, SERC, Temple University, Philadelphia, Pennsylvania 19122, USA}
\address[label5]{Department of Physics, Lebanon Valley College, Annville, Pennsylvania 17003, USA}

\begin{abstract}
We calculate energy-energy correlations for two hadrons produced inside a jet in transversely polarized proton-proton collisions.  We make numerical predictions based on a simple model that utilizes a previous global QCD analysis of dihadron fragmentation and transversity parton distribution functions.  The results show remarkable agreement with a very recent STAR measurement.  We also find the data at large jet transverse momentum have a slight preference for extractions of transversity that are consistent with lattice QCD computations of the nucleon tensor charges. Overall, this work provides further evidence for the underlying non-perturbative mechanism of near-side energy-energy correlators as well as highlights the potential for these observables to probe transverse-spin effects inside the nucleon.

\end{abstract}

\maketitle

\renewcommand*{\thefootnote}{\arabic{footnote}}
\setcounter{footnote}{0}

\section{Introduction}
Transverse single-spin asymmetries (TSSAs) and energy-energy correlators (EECs) represent two long-standing phenomena in the study of perturbative QCD and hadronic structure, both dating back to the 1970s~\cite{Klem:1976ui,Bunce:1976yb,Basham:1978bw,Basham:1978zq}.  The former allow one to explore 3-dimensional imaging of hadrons and multi-parton correlations inside of them (see the review in Ref.~\cite{Lorce:2025aqp}), while the latter provide a platform to investigate asymptotic freedom and confinement in QCD (see the review in Ref.~\cite{Moult:2025nhu}).  Over the last few years, theoretical connections have been made between these two areas~\cite{Liu:2022wop,Kang:2023big,Liu:2024kqt,Gao:2025evv,Cao:2025icu,Bhattacharya:2025bqa,Kang:2026pro}, with a very recent measurement from the STAR Collaboration at the Relativistic Heavy Ion Collider (RHIC) producing the first experimental data in this domain~\cite{STAR:2026epw}.  STAR measured both one-point (OPEC) and two-point (EEC) energy correlators inside a jet in high-energy collisions of transversely polarized and unpolarized protons, i.e., $p^\uparrow p \to ({\rm jet}, h)\,X$ and $p^\uparrow p \to ({\rm jet}, h_1 h_2)\,X$, respectively. Both observables are sensitive to the transversity parton distribution function (PDF)~\cite{Ralston:1979ys}.  This function has received particular attention since it can be used to compute the up and down quark tensor charges, $\delta u$ and $\delta d$, of the nucleon, which bridge different areas of nuclear physics, like QCD phenomenology~\cite{Anselmino:2007fs, Anselmino:2008jk, Bacchetta:2011ip, Bacchetta:2012ty, Anselmino:2013vqa, Anselmino:2015sxa, Kang:2015msa, Radici:2015mwa, Radici:2016lam, Lin:2017stx, Radici:2018iag, DAlesio:2020vtw, Cammarota:2020qcw, Gamberg:2022kdb, Cocuzza:2023oam,Cocuzza:2023vqs, Boglione:2024dal, Zeng:2024gun}, low-energy beyond the Standard Model (BSM) studies (see, e.g., Refs.~\cite{Herczeg:2001vk,Erler:2004cx,Pospelov:2005pr,Severijns:2006dr,Cirigliano:2013xha,Courtoy:2015haa,Yamanaka:2017mef,Liu:2017olr,Gonzalez-Alonso:2018omy}), lattice QCD (LQCD)~\cite{Gupta:2018qil, Gupta:2018lvp, Yamanaka:2018uud, Hasan:2019noy, Alexandrou:2019brg, Harris:2019bih, Horkel:2020hpi, Alexandrou:2021oih, Park:2021ypf, Tsuji:2022ric, Bali:2023sdi, QCDSFUKQCDCSSM:2023qlx,Gao:2023ktu, Djukanovic:2024krw, Alexandrou:2024ozj, Wang:2025nsd}, and model calculations~\cite{He:1994gz, Barone:1996un, Schweitzer:2001sr, Gamberg:2001qc, Pasquini:2005dk, Wakamatsu:2007nc, Lorce:2007fa, Yamanaka:2013zoa, Pitschmann:2014jxa, Xu:2015kta, Wang:2018kto, Liu:2019wzj,Ghim:2025gqo}.

The theoretical framework for the OPEC in $p^\uparrow p \to ({\rm jet}, h)\,X$ was established in Ref.~\cite{Gao:2025evv} and predictions were provided in the STAR paper to compare with their experimental data.  However, the same was not available for the EEC in $p^\uparrow p \to ({\rm jet}, h_1 h_2)\,X$.  In Ref.~\cite{Kang:2026pro}, which had been finalized just as the STAR paper~\cite{STAR:2026epw} became available, we developed the theoretical underpinnings, based on dihadron fragmentation~\cite{Collins:1993kq,Jaffe:1997hf,Bianconi:1999cd,Radici:2001na,Pitonyak:2023gjx,Rogers:2024nhb,Pitonyak:2025lin}, for EEC-type TSSA observables in  semi-inclusive deep-inelastic scattering (SIDIS) and $e^+e^-$ annihilation for the case where the angle $\chi$ between the two hadrons $\approx 0$ (near side).  In this Letter, we now extend our work in Ref.~\cite{Kang:2026pro} to $p^\uparrow p \to ({\rm jet}, h_1 h_2)\,X$ and generate predictions to compare with the STAR data.  Within a simple model that utilizes a previous global QCD analysis~\cite{Cocuzza:2023oam, Cocuzza:2023vqs} of dihadron fragmentation functions (DiFFs) and transversity PDFs, we find remarkable agreement with the STAR data.

The paper is organized as follows.  In Sec.~\ref{s:theory} we review the dihadron fragmentation framework for EECs and derive the leading-order (LO) analytical result for $p^\uparrow p \to ({\rm jet}, h_1 h_2)\,X$.  In Sec.~\ref{s:num} we discuss our numerical methodology, compare our theoretical predictions to the STAR data, and discuss our results.  We conclude in Sec.~\ref{s:sum}.

\section{Theoretical Background} \label{s:theory}
In Ref.~\cite{Kang:2026pro} we introduced a framework for computing transverse-spin dependent near-side EECs in SIDIS and $e^+e^-$ annihilation with the underlying mechanism being dihadron fragmentation, building on our previous work on unpolarized EECs~\cite{Kang:2025zto}.  We now extend this framework to proton-proton collisions, focusing on the process $p^\uparrow p \to ({\rm jet}, h_1 h_2)\,X$, in which the EEC is measured between the two hadrons $h_1$ and $h_2$ within the identified jet. We recall from Ref.~\cite{Kang:2026pro} the following nonperturbative EEC-DiFF functions, with $i=q$ or $g$, which enter the present calculation:
\begin{align}
\mathcal{D}^{h_1h_2/i}(z_\chi, Q^2)\equiv&\int \!\!d\xi_1 \!\int \!\!d\xi_2\int \!\!d^2\!\vec{R}_T\,\delta\!\left(z_\chi-\frac{R_T^2}{Q^2}\frac{\xi^2}{\xi_1^2\xi_2^2}\right) \xi_1\xi_2\,D_1^{h_1h_2/i}(\xi_1,\xi_2,\vec{R}_T)\,,\label{e:EEC_DiFF}\\[0.3cm]
\mathcal{H}^{h_1h_2/i}(z_\chi, Q^2)\equiv&\int \!\!d\xi_1 \!\int \!\!d\xi_2\int \!\!d^2\!\vec{R}_T\,\delta\!\left(z_\chi-\frac{R_T^2}{Q^2}\frac{\xi^2}{\xi_1^2\xi_2^2}\right) \xi_1\xi_2\,\,\frac{R_T}{M_h}H_1^{\sa \,h_1h_2/i}(\xi_1,\xi_2,\vec{R}_T)\,,\label{e:H1EEC_DiFF}
\end{align}
where  $\xi_1,\xi_2$ are the fractions of the light-cone momentum of the fragmenting quark $q$ (with momentum $k$) carried by the hadrons $h_1, h_2$, $\vec{R}_T\equiv \frac{1}{2}(\vec{P}_{1T}-\vec{P}_{2T})$ is (half of) their relative transverse momentum ($R_T\equiv|\vec{R}_T|$), $z_\chi\equiv \frac{1}{2}(1-\cos\chi)$, and $Q$ is the relevant hard scale.  For the measurement of the EEC in $p^\uparrow p \to ({\rm jet}, h_1 h_2)\,X$ by STAR~\cite{STAR:2026epw}, we have $Q^2\to P_{JT}^2e^{2\eta}$ (see Eq.~\eqref{e:z12} below), where $\eta$ and $\vec{P}_{JT}$ are the rapidity and transverse momentum, respectively, of the jet.
The total momentum of the dihadron is denoted by $P_h$, with $M_h^2=P_h^2$ its squared invariant mass, and $\xi\equiv \xi_1+\xi_2$.  We work in the ``dihadron frame'' where $P_h$ has no transverse component and a large lightcone-minus component.  
The operator definitions of the two DiFFs, $D_1^{h_1h_2/q}$ and $H_1^{\sa\, h_1h_2/q}$, that enter Eqs.~\eqref{e:EEC_DiFF}, \eqref{e:H1EEC_DiFF} are given by~\cite{Pitonyak:2023gjx} (see also Ref.~\cite{Pitonyak:2025lin})
\begin{align}
    &D_1^{h_1h_2/q}(\xi_1,\xi_2,\vec{R}_T) = \!\frac{\xi^2}{64\pi^3\xi_1\xi_2}\!\int\! \!d^2\vec{k}_T\,\sum_X\hspace{-0.5cm}\int\! \int\!\!\frac{dx^+\!d^2\vec{x}_\perp}{(2\pi)^3}e^{ik\cdot x}\,{\rm Tr}\langle 0|\gamma^-\,\psi_q(x)|P_1,P_2;X\rangle\langle P_1,P_2;X|\bar{\psi}_q(0)|0\rangle\big|_{x^-=0}\,,\\
    &\!\!\!-\frac{\epsilon_T^{ij}R_T^j}{M_h}H_1^{\sa\,h_1h_2/q}(\xi_1,\xi_2,\vec{R}_T) = \!\frac{\xi^2}{64\pi^3\xi_1\xi_2}\!\int\! \!d^2\vec{k}_T\,\sum_X\hspace{-0.5cm}\int\! \int\!\!\frac{dx^+\!d^2\vec{x}_\perp}{(2\pi)^3}e^{ik\cdot x}\,{\rm Tr}\langle 0|i\sigma^{i-}\gamma^5\,\psi_q(x)|P_1,P_2;X\rangle\langle P_1,P_2;X|\bar{\psi}_q(0)|0\rangle\big|_{x^-=0},
\end{align}
where we have suppressed gauge links, and a color average is understood. The Levi-Civita tensor is defined as $\epsilon_T^{ij}\equiv \epsilon^{-+ij}$, with $\epsilon^{0123}=+1$.

We first derive the LO differential cross section for the process
\begin{equation}
    p^\uparrow(\vec{S}_T,P) + p(P') \to \Big[{\rm jet}(P_J), h_1(P_1)\,h_2(P_2)\Big]+X\,,
\end{equation}
which will be differential in $\eta$ and $\vec{P}_{JT}$ as well as the longitudinal momentum fractions $\tau_1$ and $\tau_2\;$ of the jet momentum carried by the hadrons $h_1$ and $h_2$, respectively, and their relative transverse momentum $\vec{R}_T$.  (We use the notation $\tau_{(1,2)}$ for momentum fractions as to not cause confusion with the variable $z_\chi$.)  Starting from the cross section for $p^\uparrow p \to ({\rm jet}, h)\,X$~(see Eqs.~(1)--(4) of Ref.~\cite{Kang:2017btw}) and using the interpretation of FFs as number densities, one can make the replacements
\begin{align}
    D_1^{h/c}(\tau,\vec{j}_T)\,d\tau\,d^2\vec{j}_T&\longrightarrow D_1^{h_1h_2/c}(\tau_1,\tau_2,\vec{R}_T)\,d\tau_1d\tau_2d^2\vec{R}_T\,,\\[0.3cm] \left[\frac{\epsilon_T^{ij}j_T^j}{\tau M_h}H_1^{\perp h/c}(\tau,\vec{j}_T)\right]d\tau\,d^2\vec{j}_T &\longrightarrow \left[-\frac{\epsilon_T^{ij}R_T^j}{M_h}H_1^{\sa\,h_1h_2/c}(\tau_1,\tau_2,\vec{R}_T)\right]d\tau_1d\tau_2d^2\vec{R}_T\,,
\end{align} 
where $\vec{j}_T$ is the transverse momentum of the (single) hadron w.r.t.~the jet, and we have made use of $\vec{k}_T=-\vec{j}_T/\tau$.
Therefore, one obtains 
\begin{equation}
    \frac{d\sigma}{d\eta \,d^2\vec{P}_{JT}\,d\tau_1d\tau_2\,d^2\vec{R}_T} = F_{UU}+\sin(\phi_S-\phi_{R_T})\,F_{UT}^{\sin(\phi_S-\phi_{R_T})}\,,
\end{equation}
where
\begin{align}
    F_{UU} &= \frac{\alpha_S^2}{S}\sum_{a,b,c}\int_{x'_{min}}^1\frac{dx'}{xx'}\frac{1}{x'S+T}\,f_1^a(x)\,f_1^b(x')\,D_1^{h_1h_2/c}(\tau_1,\tau_2,\vec{R}_T)\,H_{ab\to c}^U(\hat{s},\hat{t},\hat{u})\,,\label{e:FUU}\\[0.3cm]
    F_{UT}^{\sin(\phi_S-\phi_{R_T})} &= -\frac{\alpha_S^2}{S}\sum_{a,b,c}\int_{x'_{min}}^1\frac{dx'}{xx'}\frac{1}{x'S+T}\,h_1^a(x)\,f_1^b(x')\,\frac{R_T}{M_h}\,H_1^{\sa h_1h_2/c}(\tau_1,\tau_2,\vec{R}_T)\,H_{ab\to c}^T(\hat{s},\hat{t},\hat{u})\,,\label{e:FUT}
\end{align}
with $\phi_S$ and $\phi_{R_T}$ denoting the azimuthal angles of $\vec{S}_T$ and $\vec{R}_T$, respectively.  The center-of-mass energy of the collision is given by$\sqrt{S}$, and we also have $T=-P_{JT}\!\sqrt{S}\,e^{-\eta}$, $U=-P_{JT}\!\sqrt{S}\,e^{\eta}$.  The hard factors $H^U$ and $H^T$ for the unpolarized and transversely polarized structure functions, respectively, depend on the partonic Mandelstam variables $\hat{s}=xx'S, \hat{t}=xT, \hat{u}=x'U$, where $x=-x'U/(x'S+T)$.    The explicit expressions for $H^U$ and $H^T$ can be found in Refs.~\cite{Owens:1986mp,Yuan:2007nd}, respectively (where $H^T$ is denoted as $H^{\rm Collins}$ in Refs.~\cite{Yuan:2007nd,Kang:2010zzb}).  The lower integration limit is given by $x'_{min}=x_Te^{-\eta}/(2-x_Te^{\eta})$, where $x_T=2P_{JT}/\sqrt{S}$.  The sum is over all possible partonic channels.

We now define an EEC for this reaction as 
\begin{align}
    {\rm EEC}_{pp}^{h_1h_2}&\equiv \frac{d\Sigma_{pp}^{h_1h_2}}{d\eta \, d^2\vec{P}_{JT}d\chi d\phi} 
    \equiv  \frac{\sin\chi}{2}\int \! d\tau_1d\tau_2d^2\vec{R}_T \,\tau_1\tau_2\, \frac{d\sigma}{d\eta \,d^2\vec{P}_{JT}\,d\tau_1d\tau_2\,d^2\vec{R}_T} \delta(z_\chi-z_{12})\delta(\phi-\phi_{R_T})\\[0.3cm]
    &=\frac{\sin\chi}{2}\frac{\alpha_S^2}{2\pi S}\sum_{a,b,c}\int_{x'_{min}}^1\frac{dx'}{xx'}\frac{1}{x'S+T}\,f_1^b(x')\left[f_1^a(x)\,\mathcal{D}^{h_1h_2/c}(Z)\,H_{ab\to c}^U(\hat{s},\hat{t},\hat{u})\right.\nonumber\\[-0.1cm]
    &\hspace{5.75cm}\left. -\sin(\phi_S-\phi)\,h_1^a(x)\,\mathcal{H}^{h_1h_2/c}(Z)\,H_{ab\to c}^T(\hat{s},\hat{t},\hat{u})\right]\nonumber \\
    &\equiv \Sigma_{UU} + \sin(\phi_S-\phi)\,\Sigma_{UT}\,,
\end{align}
where we have introduced the variable $z_{12}\equiv \frac{1}{2}\!\left(\!1-\tfrac{\vec{P}_1\cdot\vec{P}_2}{|\vec{P}_1||\vec{P}_2|}\right) = \frac{1}{2}(1-\cos\theta_{12})$, where $\theta_{12}$ is the angle between $h_1$ and $h_2$, and $Z\equiv z_\chi Q^2=z_\chi P_{JT}^2e^{2\eta}\;$. The scale is $Q^2=P_{JT}^2 e^{2\eta}$ rather than $Q^2=P_{JT}^2$ because STAR reports the measurement in terms of the opening angle $\chi$ between the two hadrons, instead of the rapidity-azimuth angular distance $R_L=\sqrt{(\Delta\eta)^2+(\Delta\phi)^2}$ used, for example, in Refs.~\cite{CMS:2024mlf,ALICE:2024dfl,STAR:2025jut}.  Up to power suppressed corrections $\sim 1/P_{JT}^2$, one can show 
\begin{equation}
    z_{12} = \frac{R_T^2}{P_{JT}^2e^{2\eta}}\frac{\tau^2}{\tau_1^2\tau_2^2}\,,\label{e:z12}
\end{equation}
with $\tau\equiv\tau_1+\tau_2$. Following Ref.~\cite{Kang:2026pro} we have written the EEC-DiFFs as depending on the single variable $Z=z_{\chi}P_{JT}^2e^{2\eta}$ to emphasize the close connection to {\it collinear factorization} for {\it single hadron} proton-proton observables.  The asymmetry is then defined as the ratio of the azimuthal-dependent and azimuthal-independent terms:
\begin{equation}
A_{UT,{\rm EEC},pp}^{\sin(\phi_S-\phi)} \equiv \frac{\Sigma_{UT}}{\Sigma_{UU}}\,, \label{e:AUTpp}
\end{equation}
where we implicitly assume a $\pi^+\pi^-$ final state, which was considered in the STAR analysis~\cite{STAR:2026epw} and the DiFF extractions of Refs.~\cite{Cocuzza:2023oam, Cocuzza:2023vqs}.  

\section{Numerical Predictions and Comparison to Experiment} \label{s:num}

We evaluate Eq.~\eqref{e:AUTpp} for the STAR measurement of the
near-side two-point EEC TSSA in
$p^\uparrow p \to ({\rm jet},\pi^+\,\pi^-)\,X$ at
$\sqrt{S}=200~{\rm GeV}$~\cite{STAR:2026epw}.  The non-perturbative
input is the same EEC-DiFF model used in Ref.~\cite{Kang:2026pro}, where the unpolarized EEC-DiFF is written as
\begin{equation}
\mathcal{D}^{h_1h_2/i}(Z)
=
\frac{Q^2}{32}
\int_0^1 d\xi
\int_{-1}^{1} d\zeta\,
\frac{\xi^4(1-\zeta^2)^2}{M_h}\,
D_1^{h_1h_2/i}(\xi,\zeta,M_h)
\bigg|_{M_h=\widetilde{M}_h} \, ,
\label{e:Deec_model_num}
\end{equation}
with
\begin{equation}
\widetilde{M}_h
=\left[
\frac{Z\,\xi^2(1-\zeta^2)}{4}
+
\frac{2M_1^2}{1+\zeta}
+
\frac{2M_2^2}{1-\zeta}
\right]^{1/2} .
\label{e:Mh_tilde_num}
\end{equation}
For the $\pi^+\pi^-$ final state we set $M_1=M_2=m_\pi$.  The chiral-odd
EEC-DiFF $\mathcal{H}^{h_1h_2/i}(Z)$ is obtained from
Eq.~\eqref{e:Deec_model_num} by replacing $D_1^{h_1h_2/i}$ with
$H_1^{\sa\,h_1h_2/i}$.  The transversity distribution
$h_1$, the unpolarized DiFF $D_1$, and the interference DiFF
$H_1^\sa$ are taken from the JAMDiFF global analysis
\cite{Cocuzza:2023oam,Cocuzza:2023vqs}; the unpolarized proton PDF $f_1$ is taken from
CT18NLO~\cite{Hou:2019qau,Hou:2016sho}.  The central prediction uses the
common scale choice $\mu=P_{JT}$.

The JAMDiFF extraction provides collinear DiFFs as functions of
$(\xi,M_h)$, while the relation in Eq.~\eqref{e:Deec_model_num}
requires also knowledge of their dependence on
$\zeta=(\xi_1-\xi_2)/\xi$.  We parameterize this unmeasured
dependence by the same Legendre profile used in Ref.~\cite{Kang:2026pro},
\begin{equation}
D_1(\xi,\zeta,M_h)
\simeq
D_1(\xi,M_h)\,F(\zeta),
\qquad
H_1^\sa(\xi,\zeta,M_h)
\simeq
H_1^\sa(\xi,M_h)\,F(\zeta),
\label{e:Legendre_profiles_num}
\end{equation}
with
\begin{equation}
F(\zeta)
=
\frac{1}{2}
+
a\,\zeta
+
b\,\frac{3\zeta^2-1}{2}.
\label{e:profile_num}
\end{equation}
The coefficients $(a,b)$ are sampled from the region allowed by the
positivity constraints on $D_1(\xi,\zeta,M_h)$ and
$H_1^\sa(\xi,\zeta,M_h)$, namely, $D_1(\xi,\zeta,M_h)\ge 0$ and $|H_1^\sa(\xi,\zeta,M_h)|\le D_1(\xi,\zeta,M_h)$.  (The fact that we use the same profile \eqref{e:profile_num} for $D_1$ and $H_1^\sa$ is also a consequence of these constraints.) We refer the reader to the Supplemental Material of Ref.~\cite{Kang:2026pro} for more details.  This sampling propagates the model
uncertainty associated with the unmeasured $\zeta$ profile.  {\it No parameter
is tuned to the STAR asymmetry data} reported in Ref.~\cite{STAR:2026epw}.  We emphasize that the data from STAR can eventually be utilized to help {\it extract} $h_1(x)$ and $\mathcal{H}^{\pi^+\pi^-/i}(Z)$ once the analogous measurements/re-analyses are carried out for near-side EEC azimuthal asymmetries in SIDIS and $e^+e^-$ annihilation, as we proposed in Ref.~\cite{Kang:2026pro}.

\begin{figure}[b!]
  \centering
  \includegraphics[width=0.66\textwidth]{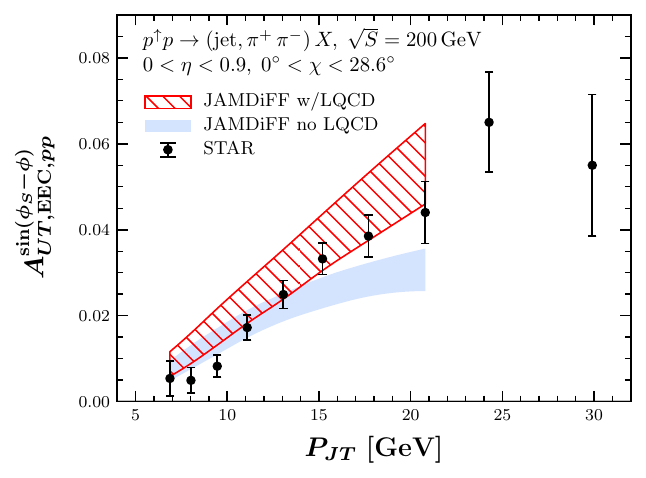}
  \caption{
  Transverse single-spin asymmetry
  $A_{UT,{\rm EEC},pp}^{\sin(\phi_S-\phi)}$ for the two-point EEC in
  $p^\uparrow p \to ({\rm jet}, \pi^+\,\pi^-)\,X$ as a function of the
   jet transverse momentum.  The points are STAR data~\cite{STAR:2026epw}, and
  the band is our prediction using Eq.~\eqref{e:AUTpp_bin} and our EEC-DiFF model described in the text.  The
  calculation is integrated over the $\eta$ and $\chi$
  ranges of the STAR measurement~\cite{STAR:2026epw}.  The theory band is displayed only for bins passing the
  kinematic cut in Eq.~\eqref{e:lambda_cut_num}. Both the input from JAMDiFF with (red, hatched) and without (light blue, solid) LQCD priors on the nucleon tensor charges are considered. Note that the STAR paper~\cite{STAR:2026epw} uses the variables $\theta_{hh}=\chi$ and $\phi_{hh}=\phi$.} 
  \label{fig:pp-eec-pt}
\end{figure}

\begin{figure}[t!]
  \centering
  \includegraphics[width=\textwidth]{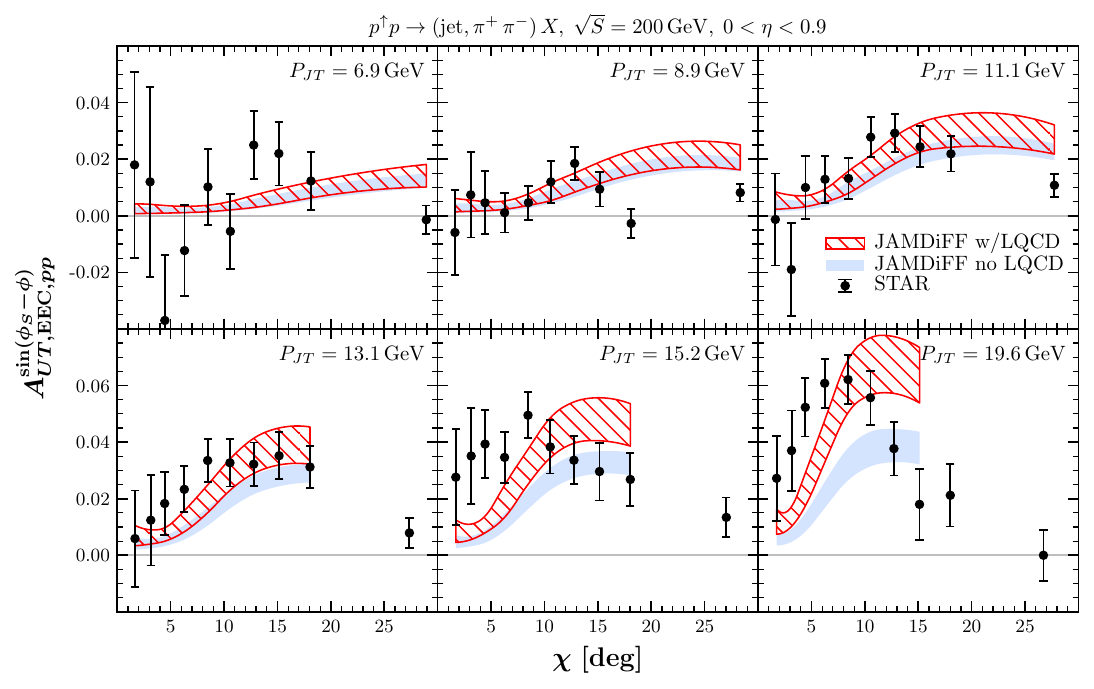}
  \caption{
  Transverse single-spin asymmetry
  $A_{UT,{\rm EEC},pp}^{\sin(\phi_S-\phi)}$ for the two-point EEC as a
  function of the hadron-pair opening angle $\chi$ in six
  jet-$P_{JT}$ bins.  The quoted $P_{JT}$ value in each panel is the
  average jet transverse momentum for that bin.  The
  points are STAR data~\cite{STAR:2026epw}, and the band is obtained from Eq.~\eqref{e:AUTpp_bin} and our EEC-DiFF model described in the text, after applying the kinematic cut in
  Eq.~\eqref{e:lambda_cut_num}. Both the input from JAMDiFF with (red, hatched) and without (light blue, solid) LQCD priors on the nucleon tensor charges are considered.  Note that the STAR paper~\cite{STAR:2026epw} uses the variables $\theta_{hh}=\chi$ and $\phi_{hh}=\phi$.}
  \label{fig:pp-eec-theta}
\end{figure}

For each STAR data point we generate a prediction for the bin-averaged observable.  For a bin
$\mathcal B$ integrated over $\eta$ and $\chi$, 
we have
\begin{equation}
A_{UT,{\rm EEC},pp}^{\sin(\phi_S-\phi)}(\mathcal B)
=
\frac{
\int_{\mathcal B} d\chi\,d\eta\,
\Sigma_{UT}
}{
\int_{\mathcal B} d\chi\,d\eta\,
\Sigma_{UU}
}\,.
\label{e:AUTpp_bin}
\end{equation}
In Fig.~\ref{fig:pp-eec-pt}, $P_{JT}$ is fixed to the published mean
value in each jet bin, and Eq.~\eqref{e:AUTpp_bin} is integrated over the
STAR ranges $0<\eta<0.9$ and $0<\chi<0.5$.  In
Fig.~\ref{fig:pp-eec-theta}, the calculation is shown as a function of
$\chi$ in six jet-$P_{JT}$ bins; in that case $P_{JT}$ and
$\chi$ are fixed to their published mean values for each point,
and the expression is integrated over $0<\eta<0.9$.

The uncertainty band is the central $68\%$ interval of the generated
prediction ensemble with 100 members.  For each ensemble member we randomly select one
JAMDiFF replica and one CT18NLO error set, draw one $(a,b)$ pair from the
allowed profile region, and evaluate the result for the three scale
choices
$\mu\in\{P_{JT}/2,\,P_{JT},\,2P_{JT}\}$.  
We only display the theory band
for kinematic regions where the argument of the EEC-DiFF satisfies following cut
\begin{equation}
\sqrt{Z_{\rm bin}} 
=
P_{JT}\,e^{\eta}\,
\sin\!\left(\frac{\chi}{2}\right)
< 4.5~{\rm GeV},
\label{e:lambda_cut_num}
\end{equation}
with $P_{JT}$, $\eta$, and $\chi$ evaluated at their bin-average
values for this purpose. This choice allows the DiFF in Eq.~\eqref{e:Deec_model_num} as one integrates to stay mostly within the $(\xi,M_h)$ range where the JAMDiFF analysis is valid.  This cut removes two data points from the
$P_{JT}$-dependent prediction and four data points from the
$\chi$-dependent prediction; the corresponding STAR points are
shown but are not used for a direct theory comparison. We mention that the largest contribution to the uncertainty comes from the JAMDiFF replicas.  The scale variation is a moderate source of uncertainty, while the choice of $(a,b)$ and CT18NLO replica provide minimal uncertainty.

The JAMDiFF analysis provides two ensembles  for the
transversity PDF and DiFFs, one obtained with and one without LQCD
constraints on the nucleon tensor charges. We generated predictions for both, labeling them ``JAMDiFF w/LQCD'' and ``JAMDiFF no LQCD''. Figure~\ref{fig:pp-eec-pt} compares these predictions with
the measured $P_{JT}$ dependence of
$A_{UT,{\rm EEC},pp}^{\sin(\phi_S-\phi)}$. Both calculations yield a
positive asymmetry that increases with the jet transverse
momentum, consistent with the trend of the STAR central values. The two
bands largely overlap at low and moderate $P_{JT}$ but become more clearly
separated at larger $P_{JT}$, where the prediction obtained with the
LQCD-constrained ensemble is generally larger and lies closer to the STAR
central values. This is due to the fact that larger $P_{JT}$ corresponds to the larger-$x$ region of the transversity PDF, where the JAMDiFF extraction with LQCD input deviates from, and becomes larger than, the extraction without LQCD input.
In addition, the no-LQCD band appears to 
flatten at lower $P_{JT}$ than suggested by
the STAR measurement, whereas the with-LQCD results maintain the upward trend displayed by the highest-$P_{JT}$ data. This behavior is obtained without introducing any
normalization or shape parameter fitted to the STAR data.

Figure~\ref{fig:pp-eec-theta} provides a more differential comparison
through the opening-angle $\chi$ dependence in fixed jet-$P_{JT}$ bins. At the
lowest jet momenta, the data fluctuate around zero with relatively large
uncertainties, while both predictions remain small and positive. As
$P_{JT}$ increases, the data and calculations develop a clear positive
asymmetry. In the intermediate $P_{JT}=8.9$, $11.1$, and
$13.1~{\rm GeV}$ bins, the predictions describe the asymmetry rise from small
$\chi$ followed by a turnover at larger opening
angles. The difference between the two JAMDiFF ensembles becomes more
pronounced in the $P_{JT}=15.2$ and $19.6~{\rm GeV}$ bins, where (in the $\chi$ region where the cut \eqref{e:lambda_cut_num} is satisfied) the
LQCD-constrained results predict a larger asymmetry and more closely follow
the measurement.  Taken together, the STAR data at large $P_{JT}$ shows a
slight preference for the predictions obtained with transversity PDFs that are consistent with LQCD computations of the nucleon tensor charges. Indeed, for the three highest $P_{JT}$ bins ($13.1, 15.2, 19.6\,{\rm GeV}$), $\chi^2/N_{\rm pts} = 2.37$ for JAMDiFF with-LQCD and $2.99$ for JAMDiFF no-LQCD, where we have used a $\chi^2$ calculation that accounts for theory uncertainty bands~\cite{Kang:2026mod}. More precise measurements at the highest $P_{JT}$ values could help further assess the compatibility between experiment and LQCD.

Overall, our analysis can account for the dominant features of the STAR proton-proton measurement~\cite{STAR:2026epw}, which is
nontrivial because the $P_{JT}$ and $\chi$ dependencies are
predicted simultaneously {\it with no refit to the STAR data}. This indicates that dihadron fragmentation can account for the underlying mechanism of near-side EECs, and the same transverse-spin effects studied previously (e.g., the transversity PDF) using ``standard'' TSSA observables can also be probed using EECs.  

\section{Summary and Outlook} \label{s:sum}
We have studied near-side EECs in transversely polarized proton-proton collisions. In particular, we considered the process
$p^\uparrow p \to ({\rm jet}, h_1 h_2)\,X$, where the EEC is measured between two hadrons inside the identified jet. Working within the dihadron fragmentation framework, we derived the corresponding factorized expression and made numerical predictions using a simple model for the relevant EEC-DiFFs.
Utilizing input from the JAMDiFF QCD global analysis~\cite{Cocuzza:2023oam, Cocuzza:2023vqs}, we find remarkable agreement with the very recent STAR measurement of this observable~\cite{STAR:2026epw}.  In addition, the STAR data at large jet transverse momentum have a slight preference for the predictions obtained with transversity PDFs that are consistent with computations of the nucleon tensor charges from lattice QCD.  This supports dihadron fragmentation as the underlying non-perturbative mechanism of near-side EECs and highlights the potential for these observables to probe transverse-spin effects inside the nucleon.  We look forward to future measurements/re-analyses of SIDIS and $e^+e^-$ annihilation data on EEC TSSAs that can be combined together in a simultaneous analysis with the proton-proton data from STAR~\cite{STAR:2026epw} to offer a simplified extraction of the transversity PDFs~\cite{Kang:2026pro}.

\section*{Acknowledgement}
We are grateful to Ting Lin for providing us with the data from the STAR measurement~\cite{STAR:2026epw}.
This work was supported by the National Science Foundation under Grants No.~PHY-2515057 (Z.K., C.Z.), No.~PHY-2412792 (A.M.), and No.~PHY-2308567, PHY-2608477 (D.P.).

\end{document}